\documentclass[prb,a4paper,10pt,twocolumn,superscriptaddress,amsmath,amsfonts,amssymb,preprintnumbers,citeautoscript]{revtex4-1}

\pdfoutput=1
\usepackage{graphicx,color,booktabs,microtype,afterpage,sidecap}
\usepackage{float}

\usepackage[colorlinks,plainpages=false,linkcolor=black,urlcolor=blue,citecolor=black,pdfpagemode=UseNone,pdfstartview=FitBH]{hyperref}
\usepackage[dvipsnames]{xcolor}
\usepackage{orcidlink}

\usepackage[T1]{fontenc}

\usepackage{amsmath}
\usepackage{lineno}

\makeatletter
\renewcommand{\@evenfoot}{\hfill\raisebox{-3em}{\bf\thepage}\hfill}
\renewcommand{\@oddfoot}{\hfill\raisebox{-3em}{\bf\thepage}\hfill}
\makeatother

\begin{document}

  \title{Imprinted atomic displacements drive spin-orbital order in a vanadate perovskite}

\author{P.~Radhakrishnan \orcidlink{0000-0002-8605-3474}}
\affiliation{Max Planck Institute for Solid State Research, Heisenbergstrasse 1, 70569 Stuttgart, Germany}

\author{K. S.~Rabinovich \orcidlink{https://orcid.org/0000-0002-2616-4623}}
\affiliation{Max Planck Institute for Solid State Research, Heisenbergstrasse 1, 70569 Stuttgart, Germany}

\author{A. V.~Boris \orcidlink{0000-0002-2062-5046}}
\affiliation{Max Planck Institute for Solid State Research, Heisenbergstrasse 1, 70569 Stuttgart, Germany}

\author{K.~F\"{u}rsich \orcidlink{0000-0003-1937-6369}}
\affiliation{Max Planck Institute for Solid State Research, Heisenbergstrasse 1, 70569 Stuttgart, Germany}

\author{M.~Minola \orcidlink{0000-0003-4084-0664}}
\affiliation{Max Planck Institute for Solid State Research, Heisenbergstrasse 1, 70569 Stuttgart, Germany}

\author{G.~Christiani}
\affiliation{Max Planck Institute for Solid State Research, Heisenbergstrasse 1, 70569 Stuttgart, Germany}

\author{G.~Logvenov}
\affiliation{Max Planck Institute for Solid State Research, Heisenbergstrasse 1, 70569 Stuttgart, Germany}

\author{B.~Keimer \orcidlink{0000-0001-5220-9023}}
\affiliation{Max Planck Institute for Solid State Research, Heisenbergstrasse 1, 70569 Stuttgart, Germany}

\author{E.~Benckiser \orcidlink{0000-0002-7638-2282}}
\email[]{E.Benckiser@fkf.mpg.de}
\affiliation{Max Planck Institute for Solid State Research, Heisenbergstrasse 1, 70569 Stuttgart, Germany}
	
\maketitle

\textbf{Perovskites with the generic composition ABO$_3$ exhibit an enormous variety of quantum states such as magnetism, orbital order, ferroelectricity and superconductivity. Their flexible and comparatively simple structure allows for facile chemical substitution and cube-on-cube combination of different compounds in atomically sharp epitaxial heterostructures. However, already in the bulk, the diverse physical properties of perovskites and their anisotropy are determined by small deviations from the ideal perovskite structure, which are difficult to control. Here we show that directional imprinting of atomic displacements in the antiferromagnetic Mott insulator YVO$_3$ is achieved by depositing epitaxial films on different facets of an isostructural substrate. These facets were chosen such that other control parameters, including strain and polarity mismatch with the overlayer, remain unchanged. We use polarized Raman scattering and spectral ellipsometry to detect signatures of staggered orbital and magnetic order, and demonstrate distinct spin-orbital ordering patterns on different facets. These observations can be attributed to the influence of specific octahedral rotation and cation displacement patterns, which are imprinted by the substrate facet, on the covalency of the bonds and the superexchange interactions in YVO$_3$. Well beyond established strain-engineering strategies, our results show that substrate-induced templating of lattice distortion patterns constitutes a powerful pathway for materials design.}

\begin{figure*}[t]
	\includegraphics[width=1.25\columnwidth]{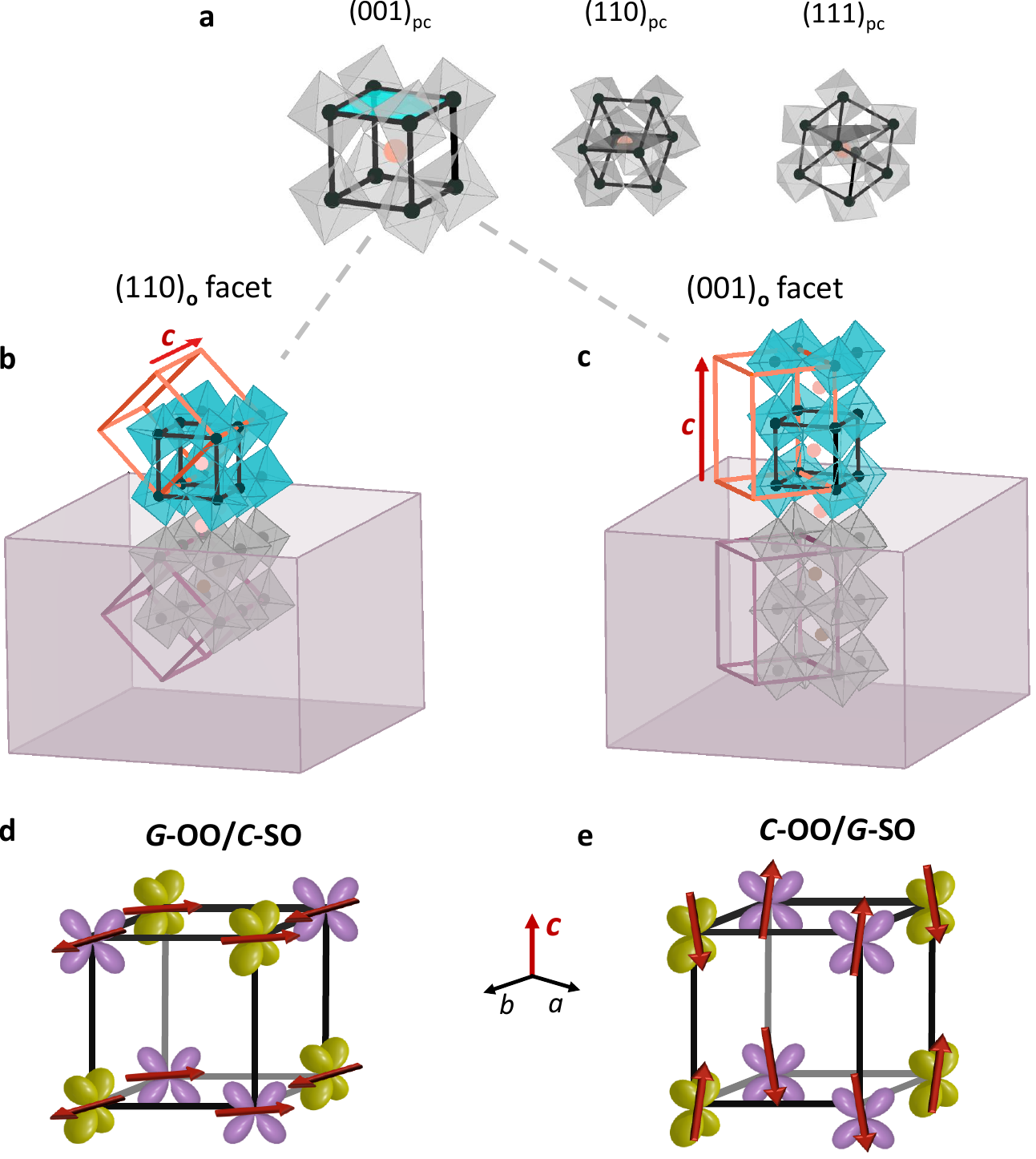}
	\caption{\footnotesize{\textbf{Different pseudocubic perovskite facets, epitaxial conditions favoring different ground states in YVO$_3$ thin films, and spin-orbital phases in bulk $R$VO$_3$}. (a) Commonly used pseudocubic (001)$_{pc}$, (110)$_{pc}$, (111)$_{pc}$ facets. The (001)$_{pc}$ facet used in this study, which corresponds to the two orthorhombic facets in (b,c), is highlighted in blue. (b, c) The epitaxial growth of YVO$_3$ thin films on substrates with orthorhombic orientation (110)$_o$ and (001)$_o$ enables the stabilization of the $G$-OO/$C$-SO and $C$-OO/$G$-SO phases, respectively. The two facets are indistinguishable in the cubic approximation and correspond to the (001)$_{pc}$ facet, as indicated by the gray dashed lines. The blue and grey octahedra represent the film and substrate, respectively. The orthorhombic and pseudocubic unit cells of the film are shown in orange and black, respectively. The film orthorhombic unit cell aligns with that of the substrate, shown in maroon. (d, e) $G$-OO/$C$-SO and $C$-OO/$G$-SO spin-orbital phases of bulk $R$VO$_3$ (adapted from Ref.~[\cite{Ren2000}]). The \textit{xz} and \textit{yz} orbitals are shown in yellow and purple and red arrows indicate the spin at each site.
}}
	\label{fig1}
\end{figure*}

Perovskites have found major applications in diverse areas including catalysis \cite{Hwang2017}, photovoltaics \cite{Huang2017}, and electronics \cite{Hwang2012}. The versatility of this structure type derives from the ability to tune both the density of mobile charge carriers and the strength and anisotropy of electron-transfer parameters between atomic sites. Exploration of the physical properties of perovskite compounds has usually proceeded by chemical substitution of bulk samples. A well-known example in the area of electronic materials is the $R$VO$_3$ family of Mott insulators ($R$ = trivalent rare earth), which exhibits different magnetic and orbital ordering patterns (some of which are associated with unusual macroscopic magnetic properties \cite{Ren1998}) upon variation of the size of the $R$ cations and consequent modulation of the vanadium-oxide bond network \cite{Miyasaka2003}. In addition, several $R$VO$_3$ compounds undergo insulator-metal transitions as mobile charge carriers are introduced by replacing $R$ by divalent species \cite{Miyasaka2000}. However, the phases accessible in bulk synthesis are limited by the number of variable parameters, such as temperature or pressure, and by structural and electronic disorder due to chemical substitution in doped variants.

Advances in the synthesis of epitaxial thin films and heterostructures have recently enabled a new set of control capabilities. For instance, interfacial charge transfer can tune the charge carrier density without introducing structural disorder \cite{Hwang2012} and substrate-induced biaxial strain may stabilize new phases through modification of bond lengths and angles \cite{Gorbenko2002, Flynn1986, Ramesh2019}. A less explored advantage of heterostructuring is the freedom to choose the crystallographic directions along which the epitaxial strain is imposed.  
Prior research in this regard, has focused on comparing films grown on different pseudocubic (pc) facets such as (001)$_{pc}$ , (110)$_{pc}$ and (111)$_{pc}$ [Fig.~\ref{fig1}(a)], that provide different surface symmetries of biaxial strain, which consequently alter the electronic properties \cite{Sando2022, Hallsteinsen2021, Grutter2010, Herranz2012, Jiagang2010}. Furthermore, the net charges of the atomic layers parallel to the substrate depend entirely on the particular pseudocubic facet [(001)$_{pc}$, (110)$_{pc}$, or (111)$_{pc}$] which can lead to distinct polarity conditions at the interface and thus to different interfacial phenomena\cite{Sando2022}. Not unlike the situation in bulk synthesis, disentangling structural and electronic effects has proven to be difficult.
Here we separate these effects and show that the atomic displacement patterns imprinted at the substrate interface alone are decisive for the stabilization of different magnetic and orbital ordering patterns in epitaxial YVO$_3$ thin films. We synthesized films on facets of a pseudocubic substrate, which are indistinguishable in the cubic approximation, have identical polarity, and almost the same lattice parameters. These facets thus differ almost exclusively in the subtle octahedral rotation and cation displacement patterns that are exposed at the substrate surface and imprinted into the films. The detection of staggered magnetic and orbital order in these films poses another set of challenges, as the x-ray diffraction and neutron scattering methods that are used for this purpose in the bulk \cite{Blake2002} are inapplicable. Therefore, we have used confocal Raman spectroscopy, which was developed for the study of thin films \cite{Hepting2014}, to investigate specific vibrational modes that are characteristic for different types of orbital order \cite{Miyasaka2006} and employed spectroscopic ellipsometry to identify optical signatures of different antiferromagnetic orders.
Remarkably, we find that different spin-orbital order patterns are realized on facets that are indistinguishable in the cubic setting. The ability to use substrate facets as templates for specific atomic displacements thus opens up interesting and potentially powerful perspectives for thin-film engineering and materials exploration.

In bulk form, the $G$-type orbital order (OO) and $C$-type spin order (SO), shown in Fig.~\ref{fig1}(d), forms the ground state of YVO$_3$. Heating above $T_{OO2/SO2}$= 77~K induces a first-order phase transition into a state with $C$-type orbital and $G$-type spin order (Fig.~\ref{fig1}(e)). Both forms of order disappear upon further heating above the N\'eel temperature at $T_{SO1}$= 116~K and the orbital order temperature at $T_{OO}$= 200~K. This sequence of transitions has been ascribed to the interplay of crystal-field effects and superexchange interactions, but many aspects of the underlying energetics remain unresolved.

\begin{figure*}[t]
	\includegraphics[width=2.0\columnwidth]{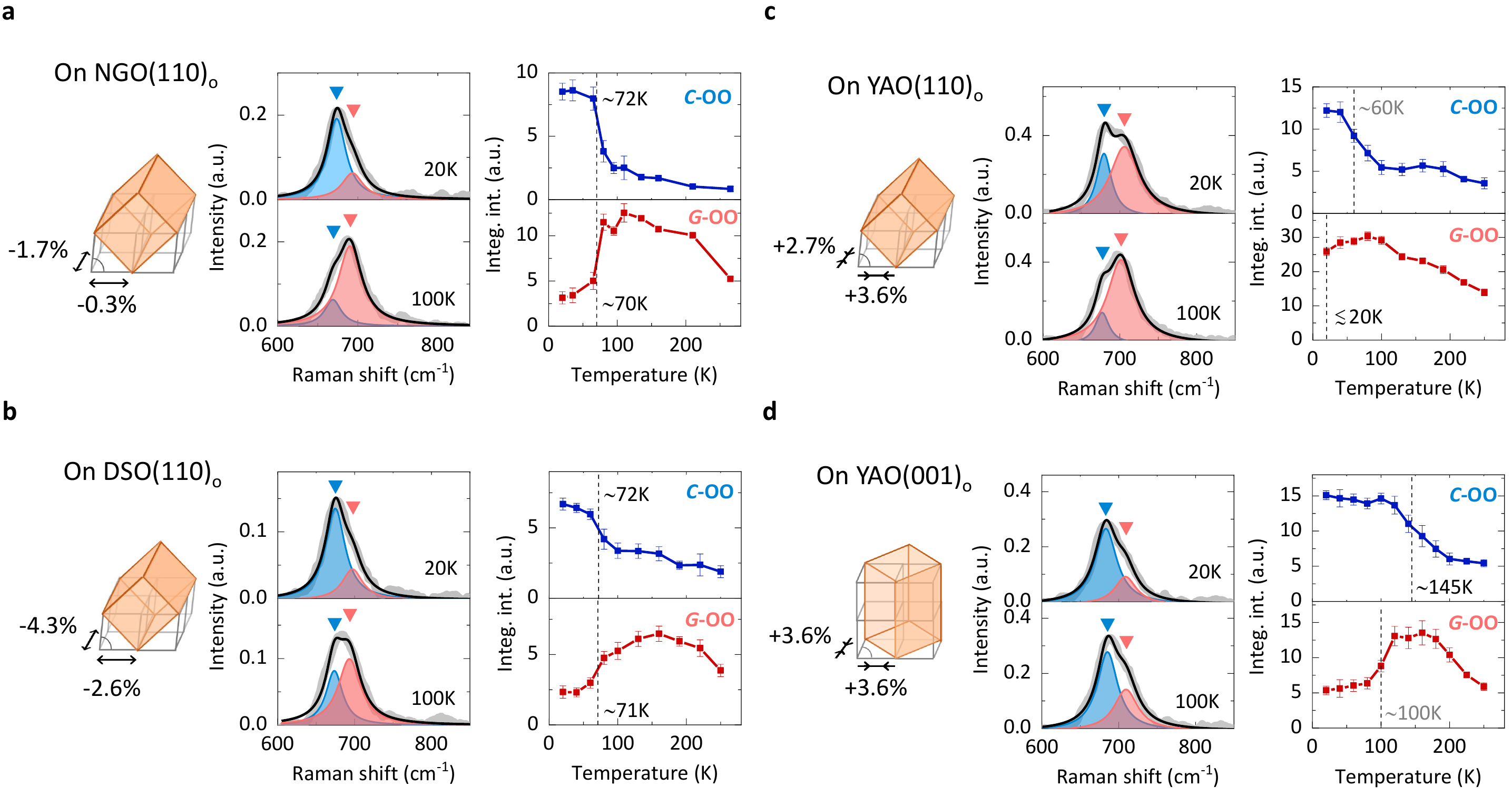}
\caption{\footnotesize{\textbf{Disentangling the effects of facet, strain and relaxation on orbital order in YVO films.} Schematic displaying the lattice mismatch of YVO$_3$ (given by (a$_l$-a$_s$)$/$a$_s$, where a$_l$ and a$_s$ are the in-plane lattice parameters of bulk YVO$_3$ and the substrate, respectively \cite{Frank1949}) (left),
Raman spectra of $B_{1g}$ mode at low and high temperatures (center), integrated intensity of $C$-OO phonon (top right) and $G$-OO phonon (bottom right) of YVO film on (a) NGO (110)$_o$, (b) YAO(110)$_o$, (c) DSO(110)$_o$ and (d) YAO(001)$_o$ substrates. The error bars denote the standard error derived from the fitting of the $C$-OO (blue triangle) and $G$-OO (pink triangle) Voight profile peaks (see Methods section). The dashed lines indicate the $T_{\text{OO2/SO2}}$ temperature of transition from the $C$-OO/$G$-SO ground state to the $G$-OO/$C$-SO intermediate temperature state. In (a, b), the $T_{\text{OO2/SO2}}$ values displayed in black and gray correspond to the fully strained and partially relaxed portion of the film, respectively (see text).  
}}\label{fig2}
\end{figure*}

We synthesized a series of YVO thin films ($\sim$ 30\,nm thickness), using pulsed laser deposition on YAlO$_3$ (YAO), NdGaO$_3$ (NGO) and DyScO$_3$ (DSO) substrates. Bulk YVO has an orthorhombic $Pbnm$ space group at room temperature, with lattice parameters $a =5.279$\,\AA, $b=5.611$\,\AA, $c= 7.572$\,\AA \cite{Reehuis2006}. All substrate materials used have the same $Pbnm$ structure. YVO is under compressive strain on YAO(110)$_o$ and YAO(001)$_o$ and under tensile strain on NGO(110)$_o$ and DSO(110)$_o$, as indicated by the lattice mismatch \cite{Frank1949} shown in the schematics in Fig.~\ref{fig2}(a-d). X-ray diffraction (XRD) revealed that in all samples the orientation of the single-domain YVO film unit cell follows that of the substrate facet, i.e., with the $c$-axis parallel to the one of the substrate. Consequently, the $c$-axis is in the plane for (110)$_o$-oriented substrates and out-of-plane for (001)$_o$-oriented substrates (see Supplementary, Section I). Further, the film on NGO(110)$_o$ was fully strained, whereas the films on YAO(110)$_o$, YAO(001)$_o$, and DSO(110)$_o$ were partially relaxed, due to the larger lattice mismatch.  

To probe the orbital order of the YVO films, we performed polarized, confocal Raman spectroscopy. Previous Raman scattering studies on bulk YVO single crystals have identified two $B_{1g}$ V-O stretching phonons around 83~meV and 86~meV that couple with the $C$-type and $G$-type OO patterns, respectively \cite{Miyasaka2006}. These modes were observed in the $y(xx)y$ polarization configuration ($x$ $\parallel$ $a$ + $b$, $y$ $\parallel$ $a$ - $b$, where $x$ and $y$ are pseudocubic directions). Fig.~\ref{fig2}(a-d) displays the Raman measurements of YVO thin films grown on NGO(110)$_o$, YAO(110)$_o$,  DSO(110)$_o$, and YAO(001)$_o$ substrates, respectively. For each sample, the $B_{1g}$ spectra and temperature dependence of the phonon mode intensities are shown on the center and right sides of the panel, respectively. The spectra at each temperature were fit with two Voigt profiles (convolutions of Gaussian and Lorentzian functions) for the $C$-OO (blue triangle) and $G$-OO (pink triangle) phonons. 

We start by comparing the films on NGO(110)$_o$ and YAO(110)$_o$ that are under tensile and compressive strain, respectively. For the film on NGO(110)$_o$ [Fig.~\ref{fig2}(a)] the integrated intensity of the $C$-OO mode is larger at low temperatures and decreases towards 100\,K, whereas the $G$-OO mode shows the opposite trend. In bulk YVO these trends across temperature are associated with the first order phase transition at $T_{\text{OO}2/\text{SO}2}$ ($\sim$ 71\,K) \cite{Miyasaka2006}.  We estimate the temperature of this transition by taking the derivative of the temperature dependence of both modes (see Supplementary, Section III for details).
This yields a value of $T_{\text{OO}2/\text{SO}2}$ $\sim$ 70-72\,K for the film on NGO(110)$_o$. The overall trends as well as $T_{\text{OO}2/\text{SO}2}$ of the film on NGO(110)$_o$ are similar to those of bulk YVO.

In the case of the film on YAO(110)$_o$ [Fig.~\ref{fig2}(b)] the temperature dependence of the $C$-OO phonon is similar to the film on NGO(110)$_o$, but has a lower value of $T_{\text{OO}2/\text{SO}2}$ $\sim$ 60\,K. The transition occurs over a broader range, which is attributed to partial relaxation. On the other hand, the temperature dependence of the $G$-OO phonon appears markedly different in comparison to the film on NGO(110)$_o$ and bulk YVO \cite{Miyasaka2006}. At the lowest temperature measured, i.e. 20\,K the intensity is high and stays relatively constant until about 100\,K, yielding $T_{\text{OO}2/\text{SO}2}$ $\sim$ 20\,K. From these results, we conclude that the $G$-OO phase is stabilized in the film on YAO(110)$_o$. Although the $T_{\text{OO}2/\text{SO}2}$ obtained from the $C$-OO and $G$-OO modes are both lower than the one of bulk YVO, there is a significant mismatch between the temperature of $G$-OO onset, and the temperature at which the $C$-OO subsides. In other words, there is a temperature range where both phases coexist. This can be understood as a consequence of the partial relaxation of the imprinted atomic displacements.
Between 20\,K and 60\,K, the layers close to the substrate that are fully coherently grown, stabilize the $G$-OO phase, whereas the increasingly relaxed layers farther from the substrate show bulk-like $C$-OO phase behavior, leading overall to the observed phase coexistence. Further, comparison of the thickness dependence of the Raman spectra of YVO films on YAO(110)$_o$ and NGO(110)$_o$ revealed that the degree of partial relaxation increases with film thickness and that the relaxed portion of the film has bulk-like behavior, which contributes to the $C$-OO peak intensity (see Supplementary, Section II).

Additionally, we investigated the effect of relaxation on the orbital ordering, to untangle its impact from the sign of strain. To this end, we synthesized a film on DSO(110)$_o$, which provides a larger magnitude of lattice mismatch than NGO(110)$_o$ and was intentionally chosen to obtain a partially relaxed film under tensile strain. The temperature dependence and transition temperatures of the $C$-OO and $G$-OO modes for this film [Fig.~\ref{fig2}(c)] are similar to the film on NGO(110)$_o$ (i.e.\ bulk-like), with a broader transition due to the partial relaxation. This means that the sign of the strain, which is the same for both YAO facets, cannot be the sole cause for the stabilization of the $G$-OO phase. Further, unlike the partially relaxed film on YAO(110)$_o$, the $T_{\text{OO2/SO2}}$ values obtained from the $C$-OO and $G$-OO modes are nearly equal for the film on DSO(110)$_o$. For the case of YVO on YAO(110)$_o$, we infer that the specific atomic displacements imprinted at the YAO(110)$_o$ substrate-film interface are the source of the $G$-OO stabilization.

Next, we examined the influence of the substrate facet, by studying a film grown on YAO(001)$_o$ [Fig.~\ref{fig2}(d)]. Our goal was to ascertain whether the effect of $G$-OO stabilization would also appear under compressive strain provided by a different facet. Following the argumentation above, we observe that the qualitative behavior of the phonons is similar to bulk YVO and the film under tensile strain on NGO(110)$_o$. However, a striking distinction is the much higher transition temperature of $T_{\text{OO}2/\text{SO}2}$ $\sim$ 145~K, obtained from the temperature dependence of the $C$-OO mode. This implies that the $C$-OO phase is enhanced in this case, in stark contrast to the film on YAO(110)$_o$, where the competing $G$-OO phase was stabilized. The temperature dependence of the $G$-OO mode also yields a larger transition temperature of $T_{\text{OO}2/\text{SO}2}$ $\sim$ 100~K compared to that of bulk YVO. 
As for the film on YAO(110)$_o$, the concurrent effects of partial relaxation and strain-induced phase stabilization create the observed difference in $T_{\text{OO}2/\text{SO}2}$. These findings imply that the substrate facet is extremely consequential in determining the properties of the YVO films, as they exhibit essentially different ground states on different facets of the same substrate material.

\begin{figure*}[t]
	\includegraphics[width=1.75\columnwidth]{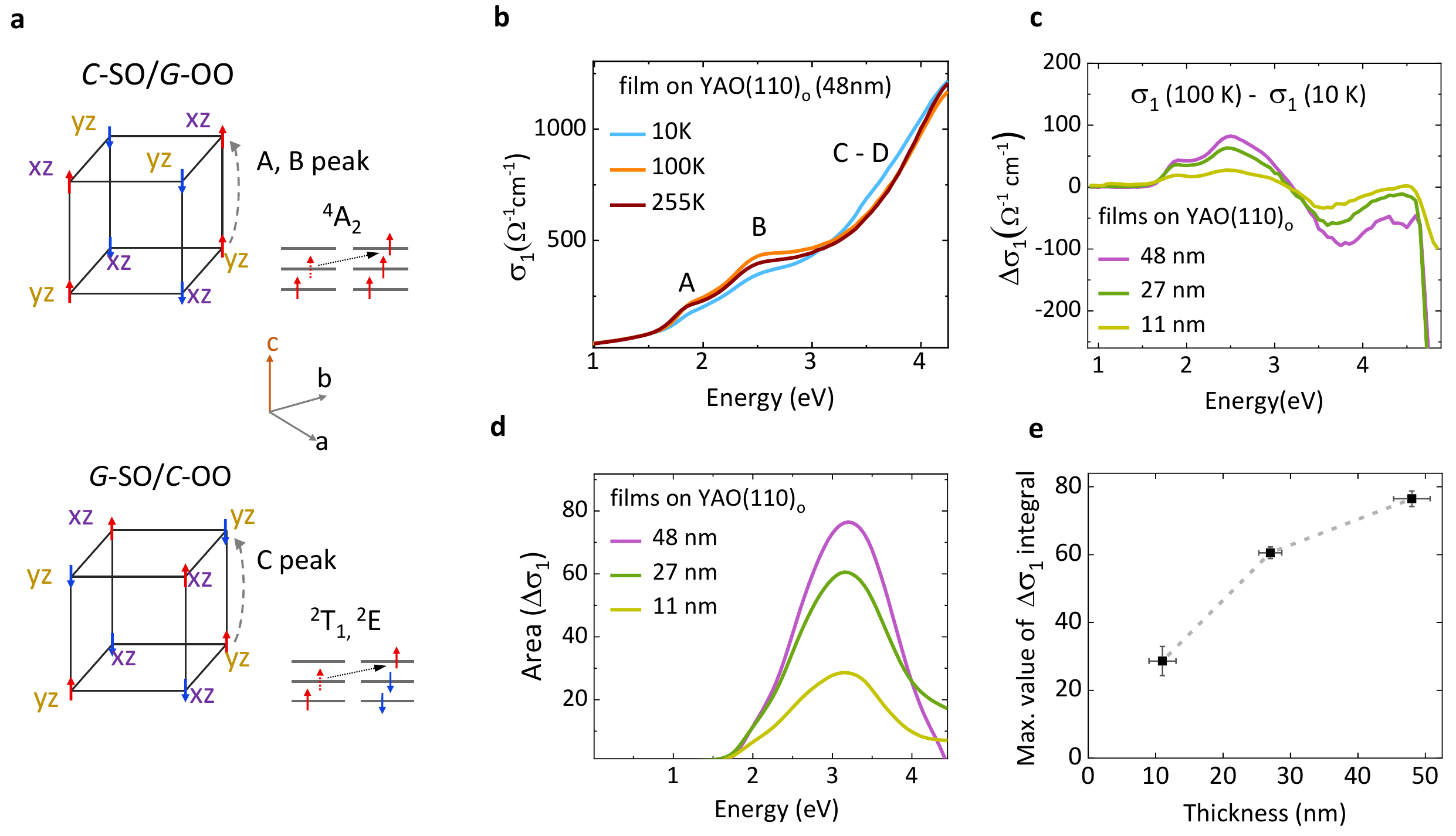}
	\caption{\footnotesize{\textbf {Electronic transitions in YVO films.} (a) Spin-orbital configurations of $C$-SO/$G$-OO and $G$-SO/$C$-OO phases and multiplet configurations of optical transitions along the $c$-axis, corresponding to peaks A, B and peak C. In bulk YVO, four multiplet peaks (A-D) were observed, where peaks A and B are associated with the high-spin ground state $^4A_2$ and peak C with the low-spin $^2E, ^2T_1$ multiplets \cite{Reul2012}. Peak D is assigned to $t_{2g}$ to $e_g$ transitions. Real part of the optical conductivity measured along $c$-axis for YVO film of 48 nm thickness on YAO(110)$_o$. (c) Difference spectra of $\sigma_1^c(100 \,K) -  \sigma_1^c(10\,K)$ for YVO films on YAO(110)$_o$ of different thicknesses. (d) Integral of $\sigma_1^c(100 \,K) -  \sigma_1^c(10\,K)$ for YVO films on YAO(110)$_o$ of different thicknesses. (e) Maximum value of $\Delta \sigma_1$ integral [curve shown in (d)] versus thickness of films on YAO(110)$_o$.}}
	\label{fig3}
\end{figure*}

Finally, to determine if the structural phase transitions indicated by the OO-coupled phonons are also associated with electronic structure changes as in bulk YVO, we used ellipsometry to measure the temperature dependent optical conductivity spectra of a YVO film on YAO(110)$_o$ (48~nm thickness). Fig.~\ref{fig3}(b)) displays the temperature dependence of the real part of the optical conductivity $\sigma_1^c$ along the $c$-axis. The positions of the multiplet peaks (A-D) matched well with those of bulk YVO, where mainly four peaks were identified \cite{Reul2012} (see caption of Fig.~\ref{fig3}). The trends of peaks A+B (high-spin multiplet) and peak C (low-spin multiplet) [Fig.~\ref{fig3}(a)] display an opposite temperature dependence (see Supplementary, Section~IV), similar to bulk YVO, which confirms that the film indeed exhibits the electronic phase transitions corresponding to the phonon behavior deduced from the Raman spectra. In particular, it is noteworthy that the optical spectra also shed light on the magnetic phase transition at $T_{\text{OO}2/\text{SO}2}$. 

To investigate the effect of epitaxy with the substrate, we compared the integral of $\sigma_{1}^c(100$~K) -  $\sigma_{1}^c(10$~K) for films of varying thicknesses grown on YAO(110)$_o$ [Fig.~\ref{fig3}(d)]. The height of this curve represents the strength of the transition at $T_{\text{OO}2/\text{SO}2}$ \cite{Kovaleva2004}. The decreasing trend [Fig.~\ref{fig3}(e)] seen as the film thickness is reduced suggests that as the film becomes thinner, this phase transition becomes increasingly suppressed. This observation is in complete accordance with the thickness dependent phonon behavior [see Fig.~\ref{fig2}(c)] and is a consequence of the $G$-OO/$C$-SO phase stabilization. Due to the gradient of imprinted atomic displacements, the fraction of the film that can contribute to the intensity reduces with the thickness of the film. Thus, the ellipsometry results compliment the conclusions drawn from Raman spectroscopy and indicate a stabilization of the $G$-OO and corresponding $C$-SO phase for YVO on YAO(110)$_o$ film. The extrapolation of the thickness dependence suggests that for films below 10~nm tickness the $C$-OO/$G$-SO phase would be on the verge of complete suppression, as the spectral weight approaches zero. This suggests that multilayers may be able to entirely stabilize this phase, where ultra-thin layers of the film can be epitaxially stabilized by sandwiching them between spacer layers \cite{Radhakrishnan2021}.

We explain the epitaxial stabilization of the spin-orbital phases based on superexchange interactions and lattice effects. The Hamiltonian of the $R$VO$_3$ system, proposed by Khaliulin \textit{et al.} \cite{Khaliullin2005} is given by ${H} = {H}_\textit{J} + {H}_\textit{V}$, and consists of two terms representing each part, respectively. For the superexchange term ${H}_\textit{J}$, considered in isolation, the lowest energy phase corresponds to a $G$-OO/$C$-SO state \cite{Khaliullin2005, Motome2003}. On the other hand, the lattice term ${H}_\textit{V}$ is a phenomenological term that supports the $C$-OO/$G$-SO phase and contains all the interactions arising from coupling to the lattice, including Jahn-Teller distortions, octahedral rotations, and $R$-O covalency effects \cite{Oles2007, Khaliullin2005, Khaliullin2004}. Theoretical studies of $R$VO$_3$, using model Hamiltonians indicated that when ${H}_\textit{J}$ dominates, the $G$-OO/$C$-SO phase is favored, whereas when ${H}_\textit{V}$ dominates, the $C$-OO/$G$-SO phase is favored \cite{Khaliullin2005}.

Let us consider the effect of compression of a lattice parameter on the superexchange interactions acting along that direction. The superexchange coupling has the form $J\propto t^2$/$U$, where $t$ is the effective $d$-$d$ (via O-2$p$) hopping amplitude and $U$ is the V-$3d$ onsite Coulomb repulsion \cite{Khomskii2014}. A decrease in lattice parameter along any direction of the perovskite structure usually produces shorter V-O bond lengths and/or smaller bond angles (i.e., angles away from 180$^\circ$) along this direction. Shorter bond lengths increase the overlap of V-3$d$ and O-2$p$ orbitals and produce larger effective $d$-$d$ hopping amplitudes, whereas smaller bond angles result in less overlap and correspondingly reduce $t$. These effects have opposite trends and the same follows analogously for expansion of the lattice parameter. Thus bond lengths and bond angles contribute oppositely to $t$ for a given change in lattice parameter. Density functional theory based studies performed for the related compounds LaVO$_3$ and YTiO$_3$ showed that the effects of the bond length change on $t$ dominate over the bond angle effect \cite{Sclauzero2016}. Therefore $c$-axis compression is expected to enhance $J$ along this axis through bond length reduction. Additionally, model Hamiltonian calculations of $R$VO$_3$ revealed that the orbital exchange constant along the $c$-axis is about ten times larger than that in the $ab$ plane for the $G$-OO/$C$-SO phase \cite{Motome2003, Khaliullin2005}. As orbital exchange is simply the orbital part of the superexchange interaction that is derived for a fixed spin order, the anisotropy of the orbital exchange constants is also transferred to the superexchange constants. Thus, any given change in bond lengths along the $c$-axis is more consequential for $J$ than those in the $ab$ plane. The enhancement of $J$ along $c$-axis therefore provides a reasonable explanation for the $G$-OO phase stabilization in (110)$_o$-oriented compressively strained films.
  
\begin{figure}[t]
	\includegraphics[width=\columnwidth]{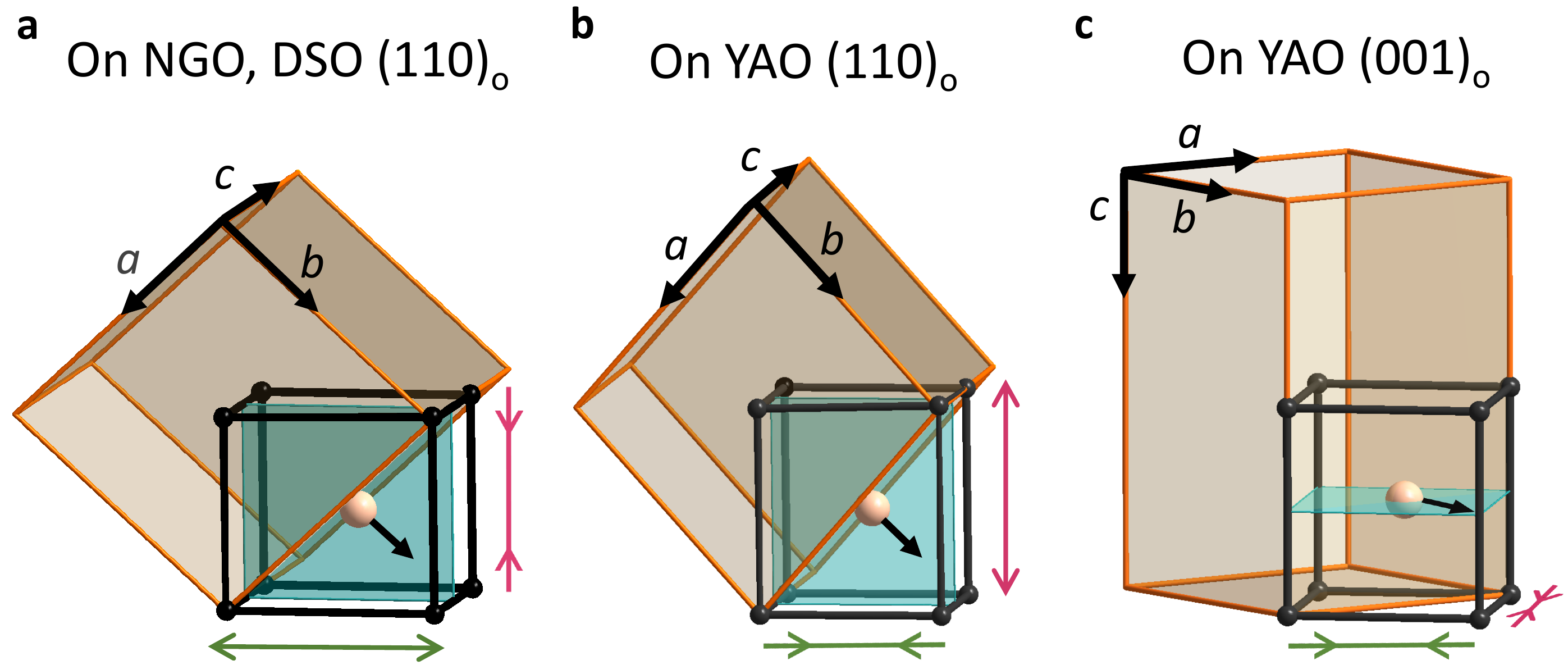}
	\caption{\footnotesize{\textbf{$A$-O covalency effect in YVO films.} Cation displacements (Y ions: orange, V ions: black) under (a) tensile strain on NGO, DSO (110)$_o$ facets,  (b) compressive strain on YAO (110)$_o$ facet, and (c), under compressive strain on YAO (001)$_o$ facet. The orthorhombic and pseudocubic unit cells are shown in orange and black, respectively. The arrow indicates the direction of the yttrium cation displacement roughly along the $b$-axis which is associated with the $d$-type JT distortion that favors the $C$-OO phase, due to increased Y-O hybridization. The Y-O distances are not expected to change appreciably in (a, b) as the lattice parameters along $a+b$ (pink arrows) decreases and $a-b$ (green arrows) increases in (a), and vice-versa in (b). On the contrary, both parameters decrease in (c) and enhance the Y-O hybridization, consequently stabilizing the $C$-OO phase in this case.}}
	\label{fig4}
\end{figure}

The lattice term ${H_V}$ is determined by several effects that together favor the $C$-OO state, which in turn stabilizes the $G$-SO state by coupling to it through Goodenough-Kanamori-Anderson rules. Of these effects, the $A$-O covalency was suggested to be a dominant contribution by Hartree-Fock calculations performed for $AM$O$_3$ perovskites \cite{Mizokawa1999}, and hydrostatic pressure studies done on single crystal YVO \cite{Bizen2008, Bizen2012}. The calculations revealed that the $d$-type Jahn-Teller (JT) distortion corresponding to the $C$-OO phase becomes lower in energy than the $a$-type JT distortion (associated with the $G$-OO phase) only when the $A$-O covalency effect is included in the calculations. 
The energy of the $d$-type distortion is lowered, because it involves $A$-cation displacements that reduce the average $A$-O nearest neighbor distances and lower the energy of the $C$-OO phase, due to increased $A$-O hybridization. Furthermore, this displacement must necessarily take place roughly along the \textit{Pbnm} $b$-axis. Fig.~\ref{fig4} depicts the $A$-O covalency effect in YVO films under epitaxial strain. For tensile (compressive) strain in the (110)$_o$ orientation, the distances along $a-b$ increase (decrease) and along $a+b$ decrease (increase), as indicated by green and pink arrows Fig.~\ref{fig4}(a, b). Therefore, in general, we would expect this term to remain relatively unchanged for this orientation. In the case of the (001)$_o$-oriented sample, the distances along both directions $a+b$ and $a-b$ decrease, leading to a net decrease in bond distances along the relevant $b$-direction, which enhances the lattice contribution that stabilizes the $C$-OO phase. The analysis of the x-ray diffraction data shows that the $b$ lattice parameter value is close to the one of bulk in the film on YAO(110)$_o$, whereas it is clearly compressed in the film on YAO(001)$_o$, as expected (see Supplementary, Sec~I). Altogether, the enhancement of lattice effects and reduction of superexchange interactions due to $c$-axis elongation provide a straightforward explanation for the $C$-OO phase stabilization under compressive strain for (001)$_o$ orientation.

In conclusion, our study on YVO$_3$ demonstrates how octahedral rotation and cation displacement patterns in perovskites can be imprinted by the choice of specific substrate facets, which allows access to different phases. We attribute the observed stabilization of different ground states exclusively to changes in the superexchange interactions and lattice effects, which are known to compete in the bulk. Since the structural distortions and the anisotropic properties studied here are typical for transition metal perovskites, the presented mechanism for material design is applicable for various epitaxial heterostructures, including titanates, chromates, and manganites. Furthermore, a similar experimental methodology to isolate the effects of the orientation of the unit cell and the sign and amplitude of strain will provide important insights into how heterostructuring modifies the electronic phases in these systems.  As such, this study broadens the perspective for a targeted manipulation of functional properties in quantum materials such as colossal magnetoresistance in manganites \cite{Yoshinori1999}, piezoelectricity in perovskite nitrites \cite{Talley2021}, tunable band gaps of transparent conducting oxides \cite{Zhang2015, Sushko2013}, and perovskite-based photovoltaics \cite{Huang2017, Zhou2014, Chen2020b}.

\bigskip

\noindent\textbf{Methods}

\begin{footnotesize}
\noindent \textbf{Synthesis and structural characterization}\\
\noindent YVO$_3$ thin films were grown by ultra-high vacuum pulsed laser deposition (UHV-PLD) from a polycrystalline target of YVO$_4$ \cite{Radhakrishnan2022} on orthorhombic substrates; YAlO$_3$(001)$_o$, YAlO$_3$(110)$_o$, NdGaO$_3$(110)$_o$, and DyScO$_3$(110)$_o$. The substrate temperature and oxygen partial pressure were optimised at about $10^{-7}$~mbar and 700$^{\circ}$C, respectively. The laser fluence and pulse repetition rates were fixed at 2~J/cm$^2$ and 5~Hz, respectively. All films were capped with LaAlO$_3$ with a thickness of about $\sim$ 4-5~nm, to preserve the V$^{3+}$ oxidation state \cite{Radhakrishnan2021}.

The films were characterized by x-ray diffraction, utilising reciprocal space mapping to determine the strain state and orientation of the films (see Supplementary, Section I). 

\vspace{2mm}

\noindent \textbf{Raman and Ellipsometry measurements}\\
\noindent Raman measurements were performed using the Jobin-Yvon LabRam HR 800 spectrometer with a HeNe laser having a wavelength of 632.8~nm. The spectra were recorded in a confocal set-up with a x100 objective and a grating with 600~grooves/mm. Laser heating effects were reduced by keeping the laser power $\sim$ 1~mW. 
For each temperature, two spectra were measured, one with the laser spot focused on the surface of the sample (raw spectra) and another with the spot focused 15~$\mu$m deeper with respect to the surface (substrate contribution), using the confocal technique \cite{Hepting2015}. The substrate signal was then subtracted from the raw spectra to obtain the Raman spectra of the film shown in Fig.~\ref{fig2}(a,b).

Consistently with previous studies on bulk $R$VO$_3$ \cite{Miyasaka2003, Miyasaka2006, Miyasaka2005}, the measurements were made with polarizations along the $x$, $y$ and $z$ pseudocubic directions, wherein $x =$ $a+b$, $y =$ $a-b$ and $c = z$ and the orthorhombic axes $a$, $b$ and $c$ are in the $Pbnm$ setting. The polarization configurations are expressed in the Porto notation: $k_{in}(\epsilon_{in},\epsilon_{out})k_{out}$, where $k$ and $\epsilon$ are the directions of propagation of light and polarization of electric field respectively with subscripts $in$ and $out$ referring to incoming and outgoing light, respectively.
The $B_{1g}$ phonons of interest were observed in the $y(xx)y$ configuration in bulk YVO. We chose to measure these phonons without an analyzer to enhance the intensity of the film spectra, i.e., in the $y(x0)y$ polarization configuration (where $0$ refers to the absence of the analyzer). Since no other phonons of $A_g$ symmetry are present in this spectral region, the removal of the analyzer has no qualitative influence on the measured spectra. Voigt profiles were used to describe the $C$-OO and $G$-OO peaks, where the parameters, area, Lorentzian width and peak positions were fitted using the Levenberg-Marquardt algorithm.

Ellipsometry measurements were performed using a variable angle spectroscopic ellipsometer (J. A. Woollam Inc.) in the energy range of 0.55 eV to 6.5 eV at an angle of incidence of 70~degrees. The measured ellipsometric parameters $\psi$ and $\Delta$ were used to obtain the complex dielectric function $\epsilon = \epsilon_1 + i \epsilon_2$ and the related optical conductivity $\sigma_1 = \omega \epsilon_2$ / $4\pi $. The ellipsometric parameters of the films were derived by fitting the data to a film-on-substrate model.
\end{footnotesize}

\bigskip

\noindent\textbf{Acknowledgements} 
\begin{footnotesize}
	\noindent We thank G.~Khaliullin for very insightful discussions and B.~Stuhlhofer for technical support with the PLD machine.
\end{footnotesize} 

\medskip

\noindent\textbf{Author contributions}
\begin{footnotesize}
    \noindent P.R.\ performed the Raman experiments, analyzed the data and conceived the project together with B.K.\ and E.B. K.F.\ and M.M.\ assisted in the Raman experiments. The spectral ellipsometry experiments were performed by K.S.R., and the data were analyzed by P.R.\ and K.S.R.\ under the supervision of A.V.B. The PLD samples were grown by P.R\ with support from G.C.\ and G.L.\ All authors contributed to the interpretation of the results.
\end{footnotesize}

%

\end{document}